\documentclass[12pt]{article}

\usepackage{amssymb}
\usepackage{subfigure}
\usepackage{color}
\usepackage{epsfig,amssymb,amsfonts,amsmath,graphicx,dsfont,cite,xfrac}
\usepackage{authblk}
\definecolor{mygray}{gray}{0.5}

\usepackage{cite}
\usepackage[colorlinks=true,linkcolor=blue,citecolor=red]{hyperref}

\newcommand{\be}{\begin{equation}}
\newcommand{\ee}{\end{equation}}
\newcommand{\bea}{\begin{eqnarray}}
\newcommand{\eea}{\end{eqnarray}}

\title{Coherent states for exactly solvable time-dependent oscillators generated by Darboux transformations}

\author[${1}$]{S. Cruz~y~Cruz}
\author[${1}$]{R. Razo}
\author[${2}$]{O. Rosas-Ortiz\thanks{Corresponding author: orosas@fis.cinvestav.mx}}
\author[${2,3}$]{K. Zelaya}

\affil[${1}$]{\footnotesize Instituto Polit\'ecnico Nacional, UPIITA, Av I.P.N 2580, C.P. 07340, M\'exico City, Mexico
}

\affil[${2}$]{\footnotesize Physics Department, Cinvestav, AP 14-740, 07000
M\'exico City, Mexico}

\affil[${3}$]{\footnotesize Centre de Recherches Math\'ematiques, Universit\'e de Montr\'eal, Montr\'eal, Qu\'ebec H3C 3J7, Canada
}
\date{}
\begin{document}

\maketitle

\begin{abstract}
The Darboux method is commonly used in the coordinate variable to produce new exactly solvable (stationary) potentials in quantum mechanics. In this work we follow a variation introduced by Bagrov, Samsonov, and Shekoyan (BSS) to include the time-variable as a parameter of the transformation. The new potentials are nonstationary and define Hamiltonians which are not integrals of motion for the system under study. We take the stationary oscillator of constant frequency to produce nonstationary oscillators, and also provide an invariant that serves to define uniquely the state of the system. In this sense our approach completes the program of the BSS method since the eigenfunctions of the invariant are an orthonormal basis for the space of solutions of the related Schr\"odinger equation. The orthonormality holds when the involved functions are evaluated at the same time. The dynamical algebra of the nonstationary oscillators is generated by properly chosen ladder operators and coincides with the Heisenberg algebra. We also construct the related coherent states and show that they form an overcomplete set that minimizes the quadratures defined by the ladder operators. These states are not invariant under time-evolution since their time-dependence relies on the basis of states and not on the complex eigenvalue that labels them. Some concrete examples are provided.

\end{abstract}


\section{Introduction}

The method introduced by Darboux in 1882 \cite{Dar82}, already 45 years before quantum mechanics was formally structured by Heisenberg, Dirac and Schr\"odinger, was addressed to the application of infinitesimal calculus in the study of surfaces \cite{Dar89}. Darboux proposed a transformation which leaves key geometric properties of certain classes of surfaces unchanged \cite{Dar82,Dar89}. For a long time the Darboux method (and its generalization developed by B\"acklund) was applied in the study of solitons \cite{Mat91}. Unexpectedly, the models introduced for the study of bosons and fermions (in the same picture)  were also associated with the Darboux transformation \cite{Mie04}. The term {\em supersymmetric} quantum mechanics came to denote the simplest case of such models and gave rise to a new branch of quantum physics which has grown stronger over the years \cite{Mie04,And04,Kha04,Suk04}. Nevertheless, the development of the Darboux method shows chronological gaps \cite{Rosu99} since the ideas that underline the transformation of surfaces developed by Darboux ``emerge, disappear and re-emerge again'' \cite{Mie04}. Such ideas find a diversity of applications in contemporary physics and mathematics \cite{Rosu99,Rosu00,Ali19,Con19a,Con19b} (see also the book \cite{Rog02}).

It is remarkable that most of the works dealing with the supersymmetric construction of exactly solvable potentials use the Darboux transformation in the spatial variable only. A notable exception is offered by the papers of Bagrov, Samsonov and Shekoyan \cite{Bag95a,Bag95b,Bag96a,Bag96b}, where a variation of the Darboux transformation is introduced to include the time-variable as a parameter. Then, one is now able to construct exactly solvable (nonstationary) time-dependent potentials as the Darboux-deformations of a given (stationary or nonstationary) potential, the solutions of which are very well known. 

The capability of solving nonstationary systems opens a diversity of applications in the trapping of particles by electromagnetic fields (see, e.g. \cite{Pau90,Gla92,Rosu96,Mih09,Mih18,Mih19}). However, the nonstationary systems are usually affected by external forces that either take energy from them or supply energy to them. That is, the corresponding Hamiltonian is not an integral of motion in such cases. No orthonormality of the basic solutions is then expected a priori, so that the determination of the observables that define uniquely the state of the system is an open problem in general. The first clue to find the appropriate invariant for this class of systems was provided by Ermakov in 1880 \cite{Erm80} (yes, a contemporary of Darboux!). Ermakov introduced a nonlinear differential equation which is connected to the Newtonian law of motion of the parametric oscillator and show that a first integral is achieved by eliminating the time-dependent frequency of oscillation from both equations. Since then, the invariant problem for time-dependent oscillators has been faced in different approaches \cite{Gla92,Lew68,Lew69,Dod75,Dod89,Dod00a,Man14,Gue15,Pad18,Gal18,Cas13,Sch13,Cru15,Cru16}, including its presence for the $x$-dependent Ermakov equation \cite{Ros15,Bla18}. Recent reports use point transformations to obtain the related invariant as a natural consequence of deforming the stationary oscillator to get nonstationary oscillators \cite{Zel19a,Zel19b}. The relevant profile of the invariant operator is that its eigenfunctions form an orthonormal basis for the space of states of the time-dependent Hamiltonian \cite{Lew68,Lew69}. Indeed, the eigenfunctions of the invariant differ from the solutions of the related Schr\"odinger equation by a time-dependent phase \cite{Lew69}.

In this work we apply the  (BSS) method introduced in  \cite{Bag95a,Bag95b,Bag96a,Bag96b} to generate nonstationary oscillators as the Darboux transformations of the well known stationary case. Our interest is twofold: 

(1) We provide a mechanism to obtain the invariant for the nonstationary oscillators so constructed, in this form the program started in Refs.~\cite{Bag95a,Bag95b,Bag96a,Bag96b} is  completed since the orthonormality of the solutions to the new time-dependent oscillators is now formally justified. We also show that the intertwining operator used in the BSS method is connected with the width of the wave-packets generated for the new oscillators, a result so far unnoticed in the literature on the matter. 

(2) We construct the ladder operators for the new oscillators and show that they close the Heisenberg algebra of the conventional oscillator. Then we show that the related coherent states are not invariant under time-evolution, in opposition to the well known behavior of the Glauber states \cite{Gla07}. The latter is because the time-dependence of the superpositions that represent the coherent states is not delivered by the complex eigenvalue which is used to label them, but by the basis of solutions itself. The coherent states so constructed minimize the quadratures associated with their ladder operators and form an overcomplete basis for the space of states of the nonstationary oscillators.

Preliminary results on the matter were already reported by two of us in \cite{Zel17}, where we used the solutions of the stationary oscillators times their phase defined by the energy to construct the states of the nonstationary oscillator (a close model has been independently developed in \cite{Con17}). Some other recent results can be consulted in, e.g. \cite{Rob18,Cen19}. Here, we construct a series of wave-packets associated to the stationary oscillator that have the profile of the Hermite-Gauss modes of quantum optics \cite{Cru17,Gre17,Gre19}. We derive the corresponding invariant (which has the profile of the Ermakov-Lewis-Reisenfeld one), and show that such states are very useful to construct the new time-dependent oscillators as well as their solutions. One of the main results included in this work is to provide the invariant operator for the nonstationary oscillators constructed via the BSS method. In this form the construction of the related coherent states is feasible as a superposition of eigenfunctions of such operator. 

The outline of the paper is as follows: In Section~\ref{Sectime}, for the sake of completeness, we revisit the BSS method. In Section~\ref{oscila} we develop our approach of the BSS method. Section~\ref{subpaq} deals with the construction of the Hermite-Gauss wave-packets for the stationary oscillator that will serve to produce the time-dependent Darboux transformation. In Section~\ref{subnon} we obtain the families of nonstationary exactly solvable oscillators. Besides we provide the dynamical algebras and coherent states involved. Section~\ref{secexamp} contains some concrete examples to show the applicability of our approach. Final discussion and conclusions are given in Section~\ref{conclu}. We provide the explicit construction of the invariants reported throughout the paper in the Appendix.

\section{Time-dependent Darboux transformation}
\label{Sectime}

Based on the Darboux method \cite{Rog02}, the approach proposed by Bagrov, Samsonov and Shekoyan (BBS) \cite{Bag95a,Bag95b,Bag96a,Bag96b} is addressed to use a time-dependent differential operator
\be
\hat L= \ell (t) \left[ \partial_x + \beta (x,t) \right], \quad \partial_x = \partial/\partial x,
\label{lop}
\ee
in order to pair the properties of two different Schr\"odinger operators,
\begin{subequations}
\be
i \hbar \partial_t - \hat H_k(x,t), \quad \partial_t = \partial/\partial t, \quad k=0,1,
\label{Sop}
\ee
\be
\hat H_k(x,t) = -\frac{\hbar^2}{2m} \partial_x^2 + V_k(x,t), \quad \partial_x^2 = \partial_x \cdot \partial_x, \quad k=0,1,
\label{Ham}
\ee
\end{subequations}
by means of the intertwining relationship
\be
\hat L \left[ i \hbar \partial_t - \hat H_0(x,t) \right] = \left[ i \hbar \partial_t - \hat H_1(x,t) \right] \hat L.
\label{intertwin}
\ee
In the previous equations it is assumed that one of the Schr\"odinger operators is exactly solvable, with very well known solutions, so the solutions of the other operator are determined via the intertwining procedure. The latter means that the subject of interest is the kernel of both Schr\"odinger operators:
\be
\left[ i \hbar \partial_t - \hat H_0(x,t) \right] \phi(x,t)=0, \quad \left[ i \hbar \partial_t - \hat H_1(x,t) \right] \psi(x,t)=0.
\label{schro0}
\ee
Clearly, the functions $\ell (t)$ and $\beta(x,t)$ introduced in (\ref{lop}) are determined such that the Schr\"odinger equations (\ref{schro0}) admit normalizable solutions. Hereafter we  assume that the solutions of the equation associated to $V_0 (x,t)$ are already known. 

The introduction of (\ref{lop}), (\ref{Sop}), and (\ref{Ham}) in (\ref{intertwin}), after some simplifications, produces the set of equations 
\begin{subequations}
\be
V_1(x,t) - V_0(x,t) = i \hbar \frac{d}{dt} \ln \ell (t) + \frac{\hbar^2}{m} \partial_x \beta(x,t),
\label{darboux1}
\ee
\be
i \hbar \partial_t \beta(x,t) + \frac{\hbar^2}{2m} \left[ \partial^2_x \beta(x,t) - \partial_x \beta^2 (x,t) \right] + \partial_x V_0(x,t)=0.
\label{riccati1}
\ee
\end{subequations}
The conventional (not time-dependent) Darboux transformation is immediately recovered from the above equations if $V_0(x,t)=V_0(x)$, for which we should make $\ell(t) = \mbox{const}$ and $\beta(x,t) = \beta(x)$. The first property that distinguishes the BBS approach from the conventional Darboux method is that a time-dependent potential $V_1(x,t)$ can be achieved even if the initial potential is a function of the position only $V_0=V_0(x)$, with the time-dependent functions $\ell(t)$ and $\beta(x,t)$ accordingly determined (see Section~\ref{oscila}). 

Similarly to the conventional case, we may introduce the additional transformation
\be
\beta(x,t) = - \partial_x  \ln u(x,t),
\label{beta}
\ee
with $u(x,t)$ a new function to be determined. Introducing (\ref{beta}) in (\ref{riccati1}) yields
\be
\left[ i \hbar \partial_t - \hat H_0(x,t)  + c_1(t)\right] u(x,t) =0,
\label{schrou}
\ee
where the time-dependent function $c_1(t)$ stems from the integration with respect to $x$. Thus, $u(x,t)$ is solution of the initial Schr\"odinger equation for which the zero point energy, represented by $c_1(t)$, is time-dependent  in general. With no loss of generality we now set $c_1(t) =0$. 

Providing a solution of (\ref{schrou}) with no zeros in $\mbox{Dom} V_0(x,t) \subseteq \mathbb R \times [t_0, \infty)$, according to (\ref{darboux1}), the new potential $V_1(x,t)$ might be a complex-valued function. In the present work we are interested in real-valued potentials, so we impose the condition
\be
\mbox{Im} \left[ i \hbar \frac{d}{dt} \ln \ell (t) + \frac{\hbar^2}{m} \partial_x \beta(x,t) \right] =0,
\ee
which is easily simplified to $\frac{d}{dt} \ln \vert \ell (t) \vert^2 = \frac{2 \hbar}{m} \mbox{Im} \partial^2_x \ln u(x,t)$. Assuming that $\ell(t)$ is real-valued we have
\be
\ell(t) = \ell_0 \exp \left\{ \frac{\hbar}{m} \int^t  d\tau \mbox{Im} \left[ \partial^2_x \ln u(x, \tau) \right] \right\},
\label{ele}
\ee
as well as the definition of the new potential
\be
V_1(x,t) = V_0 (x,t) - \frac{\hbar^2}{2m} \partial^2_x \ln \vert u(x,t) \vert^2,
\label{newV}
\ee
where $\mbox{Im} \left[ \partial^2_x \ln u(x,t) \right]$ is a constant with respect to $x$. That is, one arrives at the additional equation
\be
\partial^3_x \ln \left( \frac{u(x,t)}{u^*(x,t)} \right) =0,
\label{real}
\ee
with $z^*$ denoting the complex conjugation of $z \in \mathbb C$. The latter result is a condition that grants a real-valued potential $V_1(x,t)$ in Eq.~(\ref{newV}). As usual in the Darboux transformations, the solutions $\psi(x,t)$ of the new potential (\ref{newV}) can be obtained from the action of the intertwining operator
\be
\psi(x,t) = \hat L \phi(x,t).
\label{psi}
\ee
Additionally, the missing state 
\be
\psi_M(x,t) \propto \frac{1}{\ell(t) u^*(x,t)}
\label{missing}
\ee
must be considered since it is also a solution of Eq.~(\ref{schro0}) as well as orthogonal to all the states $\psi(x,t)$ constructed through Eq.~(\ref{psi}). Indeed, it may be shown that $\psi_M (x,t)$ satisfies the equation $\hat L^{\dagger} \psi_M=0$, from which it follows $(\psi_M, \hat L \phi) = ( \hat L^{\dagger} \psi_M, \phi)=0$.

\section{Nonstationary oscillators via the BSS approach}
\label{oscila}

Consider a stationary oscillator with constant frequency $\omega_0$, defined by the potential
\be
V_0(x,t) = V_0(x) = \frac12 m \omega_0^2 x^2.
\label{oscila1}
\ee
To obtain the intertwining operator (\ref{lop}) we have to solve the Schr\"odinger equation (\ref{schrou}) by finding a function $u(x,t)$ which is free of zeros in $\mbox{Dom} V_0(x,t) = \mathbb R \times [t_0, \infty)$. Our option is to construct a wave-packet of the oscillator (\ref{oscila1}) with the appropriate profile. Keeping this in mind, we first obtain the basic solutions $\phi(x,t)$ of the Schr\"odinger equation associated to potential (\ref{oscila1}). Then, using such a basis, we get the transformation function $u(x,t)$.

\subsection{Oscillator wave-packets}
\label{subpaq}

Following \cite{Cas13}, let us assume that the wave-packet
\be
\phi_{WP}(x,t) = N(t)  \exp \left\{ i S(t) \left[ x- \langle \hat x \rangle (t)  \right]^2  + \frac{i}{\hbar} \langle \hat p \rangle (t) \left[ x- \langle \hat x \rangle (t) \right] + i K(t)
\right\},
\label{u}
\ee
is a solution of the Schr\"odinger equation (\ref{schrou}) with $c_1(t)=0$, $\langle \hat x \rangle(t) := \eta(t)$, $\langle \hat p \rangle (t)= m  \dot\eta(t)$, and $\dot z(t) = \frac{d}{dt} z(t)$.  In turn, the mean value of position is given by
\be
\langle \hat x \rangle (t) := \langle u(t) \vert \hat x \vert u(t) \rangle = \int_{\mathbb R} dx u^*(x,t) x u(x,t).
\ee
The purely time-dependent functions $S(t)$, $K(t)$, and the normalization factor $N(t)$ are determined in the sequel.

Comparing (\ref{u}) with a conventional (normalized) wave-packet of the stationary oscillator \cite{Ros19},
\be
\Phi_{WP}(x)= \frac{1}{\left[2\pi \left( \Delta \hat x \right)^2 \right]^{1/4}} \exp \left[ -\frac{(x - \langle \hat x \rangle )^2}{4 \left( \Delta \hat x \right)^2 } + i \frac{\langle \hat p \rangle (x - \langle \hat x \rangle)}{\hbar} \right],
\label{stationary}
\ee
one realizes that the time-dependent function $S(t)$ is in general complex-valued $S(t)= S_R(t) + i S_I(t)$, where the imaginary part $S_I(t)$ should be related to the time-dependent position variance $(\Delta \hat x)^2 (t)= \langle \hat x^2 \rangle(t) - \langle \hat x \rangle^2(t)$ through $S_I(t) = \frac{1}{4 (\Delta \hat x)^2 (t)}$. Besides, the maximum of the wave-packet is located at $\eta(t) = \langle x \rangle(t)$ and thus, must follow a classical trajectory. Indeed, after substituting (\ref{u}) in (\ref{schrou}), one gets the condition for normalization 
\be
\frac{\dot N (t)}{N(t)} = - \frac{\hbar}{m} S(t),
\label{normal}
\ee
the nonlinear Riccati equation
\be
\frac{2 \hbar}{m} \dot S + \left( \frac{2 \hbar}{m} S \right)^2 + \omega_0^2=0,
\label{riccati2}
\ee
the expression for the $K$-function
\be
K(t)  =\frac{1}{2\hbar} \langle \hat p \rangle (t) \langle \hat x \rangle (t),
\ee
as well as the classical equation of motion obeyed by the maximum of the wave-packet
\be
\ddot \eta(t) + \omega_0^2 \eta(t) =0.
\label{newton}
\ee
It is a matter of substitution to show that (\ref{riccati2}) decouples into the system
\be
\frac{2 \hbar}{m} S_R(t) = \frac{\dot \alpha(t)}{\alpha(t)}, \quad S_I(t) = \frac{\lambda}{\alpha^2(t)}, \quad \lambda= \mbox{const},
\label{complex}
\ee
where $\alpha(t)$ satisfies the Ermakov equation (see \cite{Cas13} for details):
\be
\ddot \alpha(t) + \omega^2_0 \alpha (t) =  \left( \frac{2\hbar \lambda}{m} \right)^2 \frac{1}{\alpha^3(t)}.
\label{erma}
\ee
Notice that $\lambda=0$ produces the coincidence of Eq.~(\ref{erma}) with the Newtonian law of motion (\ref{newton}). Besides, it also yields $S_I=0$ in (\ref{complex}), so the wave-packet $\phi_{WP}(x,t)$ introduced in (\ref{u}) becomes a real-valued function $N(t)$ times a phase, which depends on $x$ and $t$. Thus, our approach considers $\lambda \neq 0$ in order to ensure a nontrivial function $S_I(t)$ as well as a Gaussian-like  wave-packet $\phi_{WP}(x,t)$.

To solve the Ermakov equation (\ref{erma}) one may use a pair of linearly independent solutions of (\ref{newton}), namely $\alpha_1(t) = \cos \omega_0 (t-t_0)$ and $\alpha_2(t) =\sin \omega_0 (t-t_0)$, to write \cite{Ros15}:
\be
\alpha(t) = \left\{ a \cos^2 \omega_0 (t-t_0) + b \sin 2 \omega_0 (t-t_0) + c \sin^2 \omega_0 (t-t_0) \right\}^{1/2}.
\label{solerma}
\ee
The nonnegative parameters $a,b$, and $c$, are such that
\be
b= \sqrt{a c - \left( \frac{2 \hbar \lambda}{m \omega_0} \right)^2}.
\label{cond}
\ee
Let us simplify the notation by making, without loss of generality,  $\lambda = \frac{m \omega_0}{\hbar}$. Hence $b=\sqrt{ac-4}$. On the other hand, using (\ref{complex}) in (\ref{normal}) we obtain the normalization factor
\be
N(t) = \frac{N_0}{\sqrt{\alpha(t)}} e^{-\frac{i}{2} \theta(t)},
\label{N}
\ee
with $N_0$ an integration constant which is fixed by normalization, and
\be
\theta(t) =  2\omega_0  \int^t_{t_0} \frac{1 }{\alpha^2(\tau)} d\tau =  \arctan \left\{ \tfrac12 [ (ac-4)^{1/2}+ c \tan \omega_0 (t-t_0) ] \right\}.
\label{fase1}
\ee
Therefore, the normalized wave-packet we are looking for acquires the Gaussian form
\be
\phi_{WP}(x,t) =  \left( \frac{2 m \omega_0}{\pi \hbar} \right)^{1/4} \frac{e^{-\frac{i}{2} \theta(t)}  e^{i\xi(x,t)}}{\sqrt{\alpha(t)} }  
\exp \left[- {\frac{m\omega_0}{\hbar} \left( \frac{x- \langle \hat x \rangle (t)}{\alpha(t)}  \right)^2 }\right], 
\label{u2}
\ee
where
\be
\xi(x,t) =  \frac{m}{2 \hbar}  \frac{\dot \alpha(t)}{\alpha(t)} \left[ x- \langle \hat x \rangle (t) \right]^2
+ \frac{1}{\hbar} \langle \hat p \rangle (t) \left[ x- \langle \hat x \rangle (t) \right] + \frac{1}{2\hbar} \langle \hat p \rangle (t) \langle \hat x \rangle (t).
\label{fase2}
\ee
As indicated above, the imaginary part of $S(t)$ defines the time-evolution of the width of the wave-packet. Namely, $S_I(t) = \frac{1}{4 (\Delta \hat x)^2 (t)} = \frac{\lambda}{\alpha^2(t)}$ produces the variance
\be
\left( \Delta \hat x \right)^2 (t)=  \frac{\hbar}{4m \omega_0} \left[ a \cos^2 \omega_0 (t-t_0) + b \sin 2 \omega_0 (t-t_0) + c \sin^2 \omega_0 (t-t_0) \right].
\label{width}
\ee
Then, the width oscillates with period $\tau = \frac{\pi}{\omega_0}$, and is such that
\be
\left( \Delta \hat x \right)^2 (t_n)= \left\{
\begin{array}{ll}
\left( \frac{\hbar }{4m \omega_0} \right) a, & t_n = n \tau + t_0\\[2ex]
\left(\frac{\hbar }{4m \omega_0}\right)  c, & t_n = \left( \frac{2n +1}{2} \right) \tau + t_0
\end{array}, \quad n= 0,1,2, \ldots
\right.
\ee
Assuming $a<c$ we see that the width $\left( \Delta \hat x \right)^2$ is minimum at integer multiples of the period $\tau$, while it is maximum at half-integer multiples of $\tau$ (of course, a similar configuration works if $a>c$). If $a=c$ then $\left( \Delta \hat x \right)^2(t)$ oscillates (up to the constant defining the units) between $a+b$ and $a- b$ with period $\tau = \frac{\pi}{\omega_0}$.

The expectation values $\langle \hat x \rangle$ and $\langle \hat p \rangle$ are obtained from the solutions of the Newtonian equation (\ref{newton}). As indicated above, they correspond to $\eta$ and $\dot \eta$, respectively. In short notation, it may be shown that they obey the rule
\begin{subequations}
\be
\vec \lambda(t) = R(t) \vec \lambda(t_0), \quad  \vec \lambda (t) = \left(
\begin{array}{c}
\langle \hat x \rangle (t)\\[1ex]
\langle \hat p \rangle (t)
\end{array}
\right),
\label{espera1}
\ee
where the rotation matrix
\be
R(t) = \left(
\begin{array}{cc}
\cos \omega_0 (t -t_0) &  \frac{1}{m \omega_0} \sin \omega_0 (t-t_0)\\[1ex]
-m \omega_0 \sin \omega_0 (t-t_0) & \cos \omega_0 (t -t_0) 
\end{array}
\right)
\label{espera2}
\ee
\end{subequations}
has the classical period $\tau_{osc} = \frac{2\pi}{\omega_0}$. That is, in the phase-space, the point $\vec x(t) = \vec \lambda(t)$ describes a circumference that passes through $\vec x(t_0)$ over and over as the time reaches any integer multiple of the period $\tau_{osc}$ \cite{Ros19}.

\subsubsection{Hermite-Gauss packets and their dynamical algebra}
\label{Sechg}

Introducing the variable
\be
\chi (x,t) = \left( \frac{ 2m\omega_0}{\hbar} \right)^{1/2} \left[ \frac{x- \langle \hat x \rangle (t)}{\alpha(t)} 
\label{z} \right]
\ee
one may rewrite the Gaussian wave-packet (\ref{u2}) as follows
\be
\phi_0(x,t) =  \left( \frac{2 m \omega_0}{\pi \hbar} \right)^{1/4} \frac{e^{-i \varepsilon_0 \theta(t)}  e^{i\xi(x,t)}}{\sqrt{\alpha(t)} }  e^{-\frac12 \chi^2(x,t)}, \quad \varepsilon_0 = \tfrac12.
\ee
It is immediate to recognize the resemblance with the ground state wave-function of the quantum harmonic oscillator. This wave-packet can be also compared with the fundamental (one-dimensional) off-axis Hermite-Gauss mode associated to parabolic refractive index optical media, see e.g. \cite{Gre17}. In this context, the variance $\left (\Delta \hat x \right)^2$ represents the oscillating beam width that encodes all the information of the propagation properties of the light beam along the optical axis. With this in mind we follow \cite{Cru17} and propose the set of Hermite-Gauss modes
\be
\phi_n(x,t) =   c_n \frac{e^{-i \varepsilon_n \theta(t)}  e^{i\xi(x,t)}}{\sqrt{ \alpha (t)} }   \varphi_n(\chi (x,t))
\label{soln}
\ee
as the basic solutions of the Schr\"odinger equation (\ref{schro0}) for the potential $V_0(x)$ given in (\ref{oscila1}). The constants $c_n$ must be fixed by normalization. The straightforward calculation shows that the new functions $\varphi_n(\chi)$ satisfy the (free of units) eigenvalue problem of the quantum stationary oscillator of mass and frequency both equal to 1,
\be
\left[ -\frac12 \frac{d^2}{d\chi^2}  + \frac{\chi^2}{2} - \varepsilon_n\right] \varphi_n( \chi) =0, \quad n=0,1,2,\ldots
\label{eigen1}
\ee
For $\varepsilon_n = n + \tfrac12$ the normalized solutions of (\ref{eigen1}) are well known
\be
\varphi_n(\chi) = \frac{1}{\sqrt{ n! }} \hat a^{+ n} \varphi_0(\chi), \quad \varphi_0 (\chi) = \frac{1}{\pi^{1/4}} e^{-\chi^2/2},
\label{sol1}
\ee
where
\be
\hat a^{\pm} = \tfrac{1}{\sqrt 2} \left( \mp \frac{d}{d \chi} + \chi \right),
\label{ladder1}
\ee
are the boson ladder operators $[\hat a^-, \hat a^+]= \mathbb I$, with $\mathbb I$ the identity operator, and
\be
\hat a^+ \varphi_n(\chi) = \sqrt{n+1} \varphi_{n+1}(\chi), \quad \hat a^- \varphi_n(\chi) = \sqrt{n} \varphi_{n-1}(\chi), \quad n=0,1,2,\ldots
\label{action1}
\ee
From (\ref{sol1}) and (\ref{soln}) we now introduce the operators (see details in \cite{Cru17}):
\be
\hat A^+ = e^{-i\theta(t)} e^{i\xi(x,t)} \hat a^+  e^{-i\xi(x,t)}, \quad \hat A^- = e^{i\theta(t)} e^{i\xi(x,t)} \hat a^-  e^{-i\xi(x,t)}, \quad \hat n_A= \hat A^+ \hat A^-,
\label{ladder2}
\ee
which satisfy the oscillator algebra
\be
[\hat A^-, \hat A^+]= \mathbb I, \quad [\hat n_A, \hat A^{\pm}] = \pm \hat A^{\pm},
\ee
and act on the functions $\phi_n(x,t)$ as follows
\be
\hat A^+ \phi_n(x,t) = \sqrt{n+1} \phi_{n+1} (x,t), \quad \hat A^- \phi_n(x,t) = \sqrt{n} \phi_{n-1}(x,t), \quad \hat n_A \phi_n(x,t)  = n \phi_n(x,t).
\label{action2}
\ee
In coordinate representation, it may be shown that the operators (\ref{ladder2}) are written as follows
\begin{subequations}
\be
\hat A^+ = -i e^{-i\theta(t)} \alpha(t) \left( \frac{1}{2\hbar} [\hat p - \langle \hat p \rangle(t)] - S^*(t) [ \hat x - \langle \hat x \rangle (t)] \right),
\label{Apos1}
\ee
\be
\hat A^- = i e^{i\theta(t)} \alpha(t) \left( \frac{1}{2\hbar} [\hat p - \langle \hat p \rangle(t)] - S(t) [ \hat x - \langle \hat x \rangle (t)] \right).
\label{Apos2}
\ee
\end{subequations}
Therefore, the functions $\phi_n(x,t)$ in (\ref{soln}) can be rewritten in the familiar (short) form
\begin{subequations}
\be
\phi_n(x,t) = \frac{1}{\sqrt{n!}} \hat A^{+n} \phi_0(x,t), \quad n=0,1,2,\ldots
\label{z1}
\ee
Equivalently
\be
\phi_n(x,t) =  \left( \frac{2}{\pi} \right)^{1/4} \frac{e^{-i(n+\frac12)\theta(t)} e^{i\xi(x,t)}}{\sqrt{2^n \alpha(t) n!}} e^{ - {\frac{m\omega_0}{\hbar} \left( \frac{x- \langle \hat x \rangle (t)}{\alpha(t)}  \right)^2 } } H_n  \left( \sqrt{ \frac{ 2m\omega_0}{\hbar} }  \left[ \frac{x- \langle \hat x \rangle (t)}{\alpha(t)} \right] \right),
\label{z2}
\ee
\end{subequations}
with $H_n(z)$ the Hermite Polynomials \cite{Olv10}. In the case $\langle \hat x \rangle(t) =0$ the functions $\phi_n(x,t)$ coincide with the well known Hermite-Gauss modes \cite{Cru17}. For $\langle \hat x \rangle(t) \neq 0$ this expression describes off-axis, tilted beams for which the wave vector follows a trajectory given by (\ref{espera1})-(\ref{espera2}).

On the other hand, it may be shown that the functions (\ref{soln}) are eigenfunctions of the dynamical invariant operator
\be
\hat I/I_0  = \left( \frac{\alpha}{m} \hat p - \dot \alpha \hat x \right)^2 + \left( \frac{2\omega_0}{\alpha} \right)^2 \hat x^2
 = \frac{\alpha^2}{m^2} \hat p^2 - \frac{\dot\alpha \alpha}{m}\{ \hat x, \hat p \} + \left( \dot \alpha^2 + \frac{4 \omega_0^2}{\alpha^2} \right) \hat x^2,
\label{invariante1}
\ee
which can be obtained by eliminating the frequency $\omega_0$ from the Newton equation of motion (\ref{newton}) and the Ermakov equation (\ref{erma}), just as it was shown by Ermakov \cite{Erm80} (see the discussion on the matter and further details in \cite{Cas13}). The constant $I_0$ has been introduced to provide the operator $\hat I$ with dimensions of action. The invariant problem for time-dependent oscillators has been faced in different approaches \cite{Cas13,Sch13,Cru15,Cru16,Lew68,Lew69,Dod75,Dod89,Gla92,Dod00a,Man14,Gue15,Pad18,Gal18}, including its presence for the $x$-dependent Ermakov equation \cite{Ros15,Bla18}. A more general treatment considers point transformations for which the related invariant arises as a natural consequence of deforming the stationary oscillator to get nonstationary oscillators \cite{Zel19a,Zel19b}. In Appendix~\ref{ApA} we offer an alternative derivation of the invariant (\ref{invariante1}). The relevant point here is that the functions (\ref{soln}), being eigenfunctions of the invariant operator $\hat I$, satisfy the orthonormality condition 
\be
\int_{\mathbb R} dx \phi_n(x,t) \phi_m^*(x,t) = \delta_{n,m},
\label{product}
\ee
which holds when the involved functions are evaluated at the same time (otherwise the orthogonality is not granted). Thus, the set $\phi_n(x,t)$ forms a complete basis for the normalizable solutions of the Schr\"odinger equation (\ref{schro0}) defined by the potential $V_0(x,t)$ we are dealing with.

\subsection{Nonstationary oscillators}
\label{subnon}

Following \cite{Zel17}, for the transformation function we write
\be
u(x,t) =  \frac{e^{-i \varepsilon \theta(t)}  e^{i\xi(x,t)}}{ \sqrt{\alpha(t)} } e^{-\frac12 \chi^2(x,t)} F(\chi(x,t)).
\label{f}
\ee
We look for a real-valued function $e^{-\frac12 \chi^2} F(\chi)$ that is a solution of the eigenvalue equation (\ref{eigen1}), with no zeros in $\mathbb R \times [t_0, \infty)$. Let us consider the general solution
\be
F(\chi(x,t)) = k_a {}_1 F_1 \left( \frac14 (1 -2 \varepsilon), \frac12, \chi^2 \right) + k_b \chi {}_1 F_1 \left( \frac14 (3 -2 \varepsilon), \frac32, \chi^2\right),
\ee
where the constants $k_a$, $k_b$, and $\varepsilon$ are to be determined. Indeed, assuming that $F$ fulfills our requirements, it is a matter of substitution to show that the condition (\ref{real}) is satisfied. Therefore, the potential (\ref{newV}) is real-valued and acquires the form
\be
V_1(x,t) = \frac12 m \omega_0^2 x^2 - \frac{\hbar^2}{m} \partial_x^2 \left[ \ln F(\chi(x,t))  \right] + \frac{2 \hbar \omega_0}{\alpha^2(t)}.
\label{vtemp}
\ee

Remark that the frequency of $V_1(x,t)$ is exactly the same as the constant frequency $\omega_0$ of the stationary oscillator (\ref{oscila1}). That is, the time-dependence of the new potential (\ref{vtemp})  arises from the additive term included by the Darboux transformation. In this respect, the nonstationary oscillators represented by such a potential increases the number of exactly solvable time-dependent oscillators already reported in the  literature, where it is usual to find oscillators with time-dependent frequency that are acted by a driving force which also depends on time. The time-dependent term included in (\ref{vtemp}) by the Darboux transformation would represent external forces that either take energy from the oscillator or supply energy to it. That is, depending on the functions $\alpha(t)$ and $F(\chi(x,t))$, we are facing a nonconservative system which has no solutions with the property of being orthogonal if they are evaluated at different times, similarly to the inner product of the Hermite-Gauss modes (\ref{product}). We give full details of the related solutions and their properties in Section~\ref{secsol2}. Notice also that the term containing the function $\alpha$ in (\ref{vtemp}) would represent a time-dependent zero point energy which may be omitted (at the cost of producing just the difference of a global phase in the solutions, see e.g. \cite{Zel19b}).

In turn, the $\beta$-function (\ref{beta}) acquires the form
\be
\beta(x,t) = -\frac{i}{\hbar} \langle \hat p \rangle (t) - 2i S(t) [x - \langle \hat x \rangle (t)] - \partial_x \left[ \ln F(\chi(x,t))  \right],
\label{beta2}
\ee
while the $\ell$-function (\ref{ele}) is simply $\ell(t) = \alpha(t)$. Recalling the expression that connects the width (variance) of the wave-packet (\ref{u2}) with the $\alpha$-function (\ref{width}), we immediately realize that the function $\ell(t)$ is associated to the standard deviation $\sqrt{\left( \Delta \hat x \right)^2 (t)}$ of $\phi_{WP}(x,t) \equiv \phi_0(x,t)$ as follows
\be
\ell(t) = \alpha(t) = \sqrt{\frac{4m\omega_0}{\hbar} \left( \Delta \hat x \right)^2 (t)}.
\label{ele2}
\ee
The derivation of the intertwining operator (\ref{lop}) is immediate by using (\ref{beta2}) and (\ref{ele2}). However, it is profitable to rewrite $\hat L$ in terms of the annihilation operator $\hat A^-$ introduced in (\ref{ladder2}). The straightforward calculation gives
\be
\hat L= - \alpha(t) \partial_x \left[ \ln F(\chi(x,t))  \right] + 2 e^{-i \theta(t)} \hat A^- ,
\label{lop2}
\ee
where we have used (\ref{Apos2}).

\subsubsection{Solutions and dynamical algebra for the time-dependent oscillators}
\label{secsol2}

The construction of the solutions to the Schr\"odinger equation defined by the nonstationary potential (\ref{vtemp}) is given by the transformation
\be
\begin{aligned}
\psi_{n+1} (x,t) & = \hat L \phi_n(x,t)\\
& = - \alpha(t) \partial_x \left[ \ln F(\chi(x,t))  \right] \phi_n(x,t) + 2 e^{-i \theta(t)} \sqrt{n} \phi_{n-1}(x,t), \quad n=0,1,2,\ldots
\end{aligned}
\label{nuevas}
\ee
The appropriate transformation function $u(x,t)$ defines a square-integrable missing state (\ref{missing}), which must be added to the solutions. We may consider $\varepsilon < \frac12$ to write $\psi_M(x,t) \equiv \psi_0(x,t)$. 

To get more insights about the new set of functions (\ref{nuevas}) we have to emphasize that they are not eigenfunctions of the Hamiltonian defined by the time-dependent potential (\ref{vtemp}). The reason is that such a Hamiltonian is not an integral of motion. Thereby, it is necessary to determine the first integral(s) that may serve as observable(s) to define uniquely the new oscillators. As indicated above, the existence of an invariant for the time-dependent oscillators was mathematically shown by Ermakov \cite{Erm80}. In the present case, the straightforward calculation shows that the functions $\psi(x,t)$ introduced in (\ref{nuevas}) are eigenfunctions of the invariant operator 
\be
\hat I_G = I_0 \left[ \frac{\alpha^2}{m^2} \hat p^2 - \frac{\dot\alpha \alpha}{m}\{ \hat x, \hat p \} + \left( \dot \alpha^2 + \frac{4 \omega_0^2}{\alpha^2} \right) \hat x^2 -\frac{2\hbar^2 \alpha^2}{m^2} \hat G(t) \right],
\label{invariante2}
\ee
where the constant $I_0$ is the same as the one introduced in (\ref{invariante1}), see the Appendix for the detailed derivation. The operator $\hat G(t)$, defined in Eq.~(\ref{eq:INV5}) of the Appendix, corresponds to the additive time-dependent term of $V_1(x,t)$. Notice that turning $\hat G(t)$ off both invariants coincide $\hat I_{G=0} = \hat I$, as expected.

Let us complete our program by following \cite{Bag95a,Bag95b} to introduce an additional operator $\hat M$, which is assumed to act on the space of states of the potential $V_1(x,t)$, as follows
\be
\hat M \psi(x,t) = \phi(x,t).
\ee
Using (\ref{nuevas}) one gets $\hat M( \hat L \phi) = \phi$. Then $\hat M \hat L = \mathbb I$, which means that $\hat M$ reverts the action of $\hat L$. The latter is significative since we can construct a new pair of operators 
\be
\hat B^{\pm} = \hat L \hat A^{\pm} \hat M,
\ee
such that
\be
\hat B^+ \psi_n(x,t) = \sqrt{n+1} \psi_{n+1}(x,t), \quad \hat B^- \psi_n(x,t) = \sqrt{n} \psi_{n-1}(x,t).
\ee
That is, $\hat B^{\pm}$ are the ladder operators in the space of states of the new potential. Indeed, the straightforward calculation yields the oscillator algebra
\be
[\hat B^-, \hat B^+]= \mathbb I, \quad [\hat n_B, \hat B^{\pm} ] = \pm \hat B^{\pm}, \quad \hat n_B = \hat B^+ \hat B^-.
\ee
Therefore, we can write
\be
\psi_n(x,t) = \frac{1}{\sqrt{n!}} \hat B^{+n} \phi_0(x,t), \quad n=0,1,2,\ldots
\label{newsol}
\ee

\subsection{Coherent states}

The bare essentials of coherent states can be expressed as a linear superposition
\be
\vert z_{CS} \rangle = \sum_{n \in {\cal I}} f_n(z) \vert \gamma_n \rangle, \quad z \in \mathbb C,
\label{super}
\ee
where the vectors $\vert \gamma_n \rangle$ generate a (separable) Hilbert space ${\cal H}$, ${\cal I} \subset \mathbb Z$ is an appropriate set of indices, and $f_n(z)$ is a set of analytical functions permitting normalization \cite{Ros19}. The superpositions (\ref{super}) satisfy some specific properties that are requested on demand. 

The relevance of the Hermite-Gauss modes introduced in Section~\ref{Sechg} for the stationary oscillator (\ref{oscila}), and the solutions introduced in Section~\ref{secsol2} for the nonstationary oscillators (\ref{vtemp}), is that both of them form orthogonal basis for their respective spaces of states. The latter, we insist, since they are eigenfunctions of the invariant operators $\hat I$ and $\hat I_G$, respectively. Therefore, we are able to construct time-dependent `coherent' superpositions (\ref{super}) for either the stationary or the nonstationary oscillators discussed in the previous sections. Additionally, we have constructed ladder operators for both systems that satisfy the oscillator algebra, so we have at hand either the algebraic (Barut-Girardello) or the group (Perelomov-Gilmore) approaches to define the coefficients $f_n$ in the conventional form. Thus, our coherent states mimic the Glauber states \cite{Gla07}; notwithstanding, their time-dependent behavior is quite different.

Let us start with the Hermite-Gauss modes, it is simple to show that the superposition
\be
\phi_z(x,t) = e^{-\frac12 \vert z \vert^2} \sum_{n=0}^{\infty} \frac{z^n}{\sqrt{n!}} \phi_n(x,t), \quad z \in \mathbb C,
\label{cs1}
\ee
is eigenvector of the operator $\hat A^-$ with eigenvalue $z$. The probabilities $\vert \phi_z(x,t) \vert^2$ follow the Poisson distribution and are not time-dependent (see \cite{Zel19c} and compare with \cite{Una18}). By construction, the state $\phi_z(x,t)$ minimizes the uncertainty associated to the quadratures
\be
\hat q_A = \frac{1}{\sqrt 2} (\hat A^+ + \hat A^- ), \quad \hat p_A = \frac{i}{\sqrt 2} (\hat A^+ - \hat A^- ), \quad [\hat q_A, \hat p_A ]= i.
\ee
Besides, from (\ref{z1}) it can be verified that
\be
\phi_z(x,t) = e^{-\frac12 \vert z \vert^2} e^{z \hat A^+} \phi_0(x,t) \equiv \hat D_A(z) \phi_0(x,t),
\ee
where we have used the fact that $\phi_0(x,t)$ is annihilated by $\hat A^-$ as well as the conventional disentangling formulae for the Heisenberg-Weyl group \cite{Ros19}. Thus, the coherent sates (\ref{cs1}) are also displaced versions (in the complex $z$-plane) of the fiducial state $\phi_0(x,t)$, which is a time-dependent function through the function $\chi(x,t)$ defined in Eq.~(\ref{z}). Moreover, using (\ref{product}) it can be verified that the set $\phi_z(x,t)$ is overcomplete, just as this occurs for the Glauber states.

To construct a first class of coherent states for the nonstationary oscillators (\ref{vtemp}) we can use the action of the operator $\hat L$ introduced in (\ref{lop2}). We immediately obtain
\be
\psi_z(x,t)= \hat L \phi_z(x,t)= -  \phi_z(x,t) \alpha(t) \partial_x \left[ \ln F(\chi(x,t))  \right] + 2 z e^{-i \theta(t)} \phi_z(x,t).
\label{cs2}
\ee
As we can see, in contrast with the conventional coherent states, neither $\phi_z(x,t)$ nor $\psi_z(x,t)$ preserve their form when they evolve in time. This is because their time-dependence is not focused on the complex eigenvalue $z$, but in the bases of states that are used to construct the superpositions.

Additionally, we can construct the eigenfunctions of the operator $\hat B^-$, which yields
\be
\widetilde\psi_z(x,t) = \hat D_B(z) \psi_0(x,t) = e^{-\frac12 \vert z \vert^2} \sum_{n=0}^{\infty} \frac{z^n}{\sqrt{n!}} \psi_n(x,t), \quad z \in \mathbb C,
\label{cs3}
\ee
where we have used, again, the conventional disentangling formulae for the Heisenberg-Weyl group. The latter states minimize the uncertainty of their respective quadratures
\be
\hat q_B = \frac{1}{\sqrt 2} (\hat B^+ + \hat B^- ), \quad \hat p_B = \frac{i}{\sqrt 2} (\hat B^+ - \hat B^- ), \quad [\hat q_B, \hat p_B ]= i,
\ee
and their probabilities are given out according with the (not time-dependent) Poisson distribution. It is clear that the coherent states $\widetilde \psi_z (x,t)$ are not invariant under time evolution  (the time-dependence is not defined by the complex eigenvalue $z$, as in the previous cases). Nevertheless, they form an overcomplete set in the space of states of the nonstationary oscillators $V_1(x,t)$ defined in (\ref{vtemp}).

\section{Examples and discussion of results}
\label{secexamp}

Next we provide some specific examples to show the applicability of our method. We have selected representative cases for the nonstationary oscillators $V_1(x,t)$ as well as for the related solutions $\psi_n(x,t)$ and coherent states $\psi_z(x,t)$. As indicated above, we shall take $\varepsilon <\frac12$ in order to get well defined missing states $\psi_M(x,t) \equiv \psi_0(x,t)$, however such a selection does not limit our approach since the oscillation theorems that apply for stationary Hamiltonians are not directly valid in the present case. Additionally, recall that $\psi_M(x,t)$ is orthogonal to any state $\psi(x,t)$ constructed through Eq.~(\ref{nuevas}), no matter the value of $\varepsilon$. So that any $\varepsilon \geq \frac12$ producing normalizable states $\psi_M(x,t)$ may be included in the set of solutions. Results in this direction will be reported elsewhere.

\subsection{Case $\varepsilon = -\frac12$}

\begin{figure}[htb]

\centering
\subfigure[~$t=0.2$  ]{\includegraphics[width=0.3\textwidth]{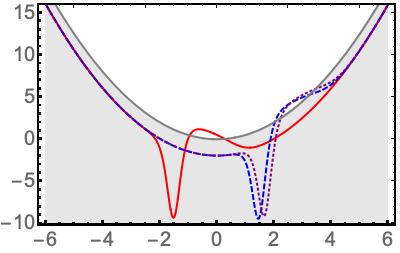}}
\hskip3ex
\subfigure[~$t=6$ ]{\includegraphics[width=0.3\textwidth]{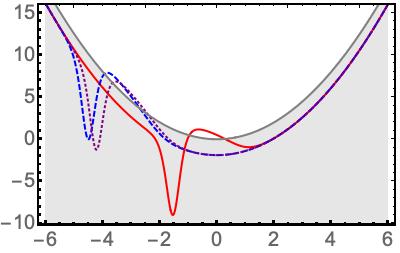}}

\caption{\footnotesize 
Nonstationary oscillators $V_1(x,t)$ defined in Eq.~(\ref{V1-Ex2}) evaluated at two different times (time is used in arbitrary units). The gray filling is a reference of the stationary oscillator (\ref{oscila1}). In all cases $t_0=0$, $m=1$, $\omega_0 =0.5$ (also in arbitrary units), and $k_a = 0.89k_b$. The curves in red, dashed-blue, and dotted-purple correspond to oscillators that departured from the initial point $(\langle \hat x \rangle_0, \langle \hat p \rangle_0)$ defined by $(0,0)$, $(3,0)$, and $(3,1)$, respectively. We have used $\alpha(t)$ with $a=1$ and $c=4$.
}
\label{figpot}
\end{figure}

\begin{figure}[htb]

\centering
\subfigure[$(0,0)$  ]{\includegraphics[width=0.3\textwidth]{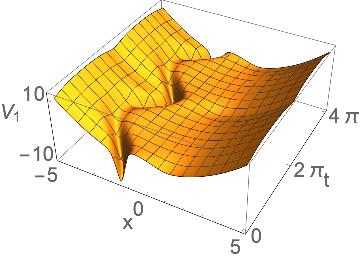}}
\hskip1ex
\subfigure[$(3,0)$ ]{\includegraphics[width=0.3\textwidth]{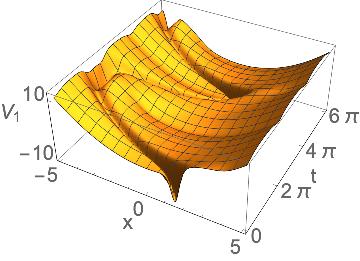}}
\hskip1ex
\subfigure[ $(3,1)$ ]{\includegraphics[width=0.3\textwidth]{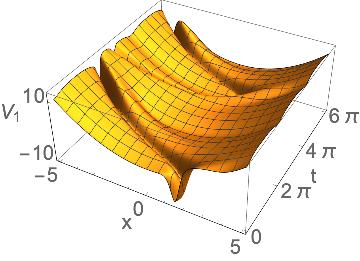}}

\caption{\footnotesize 
Time-evolution of the nonstationary oscillators $V_1(x,t)$ shown in Figure~\ref{figpot} for the indicated  values of the initial points $(\langle \hat x \rangle_0, \langle \hat p \rangle_0)$.
}
\label{PotV1}
\end{figure}

For $\varepsilon = -\frac12$ the function (\ref{f}) becomes
\be
F(\chi) = e^{\chi^2} \left[k_a + \frac{\sqrt{\pi}}2 k_b \; \mathrm{Erf}(\chi) \right],
\ee
so that the operator $\hat L$ and potential $V_1$ are respectively given by
\begin{subequations}
\be
\label{L-Ex2}
\hat L = - 2\left(\frac{2m\omega_0}\hbar\right)^{1/2} \left[\frac{k_b e^{-\chi^2(x,t)}}{2k_a+\sqrt{\pi}k_b \; \mathrm{Erf}\left(\chi(x,t)\right)} +  \chi(x,t)\right]
+ 2 e^{-i\theta(t)} A^-
\ee
and
\be
\label{V1-Ex2}
V_1(x,t) = \frac12 m \omega_0^2 x^2 - \frac{2\hbar^2}{m\alpha(t)} \left(\frac{2m\omega_0}{\hbar}\right)^{1/2} \partial_x \left[\frac{k_b e^{-\chi^2(x,t)}}{2k_a + \sqrt{\pi} k_b \; \mathrm{Erf}(\chi(x,t))} \right]  - \frac{2\hbar \omega_0}{\alpha^2(t)}.
\ee
\end{subequations}
To avoid singularities in $V_1(x,t)$ we take $\vert k_a \vert > \frac{\sqrt{\pi}}2 \vert k_b \vert$. Figure~\ref{figpot} shows the behavior of these potentials at two different times (measured in arbitrary units). We can identify a local `deformation' of these potentials with respect to the stationary oscillator $V_0(x)$ defined in (\ref{oscila1}). Such a perturbation oscillates around its initial position along the parabola described by the stationary oscillator. The latter is better appreciated in Figure~\ref{PotV1}, where the time-evolution of the nonstationary oscillators $V_1(x,t)$ is depicted for large enough time intervals.

\begin{figure}[htb]

\centering
\subfigure[$(0,0)$, $\vert \psi_0 (x,t)\vert^2$  ]{\includegraphics[width=0.3\textwidth]{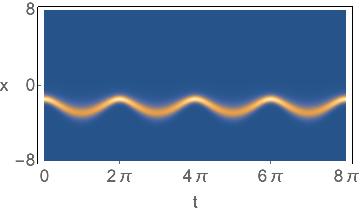}}
\hskip1ex
\subfigure[$(0,0)$, $\vert \psi_1 (x,t)\vert^2$ ]{\includegraphics[width=0.3\textwidth]{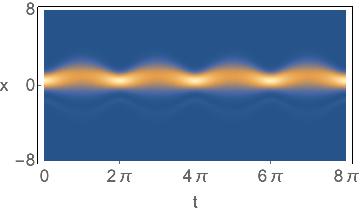}}
\hskip1ex
\subfigure[ $(0,0)$, $\vert \psi_2 (x,t)\vert^2$  ]{\includegraphics[width=0.3\textwidth]{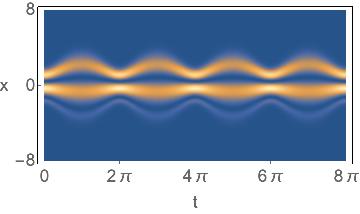}}

\vskip1ex
\subfigure[$(3,0)$, $\vert \psi_0 (x,t)\vert^2$  ]{\includegraphics[width=0.3\textwidth]{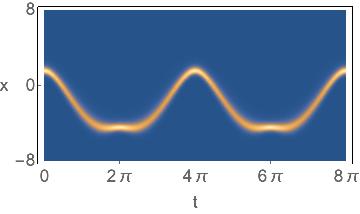}}
\hskip1ex
\subfigure[$(3,0)$, $\vert \psi_1 (x,t)\vert^2$ ]{\includegraphics[width=0.3\textwidth]{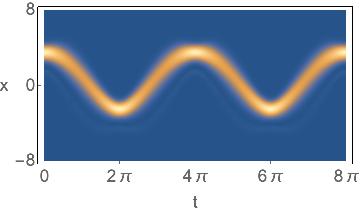}}
\hskip1ex
\subfigure[ $(3,0)$, $\vert \psi_2 (x,t)\vert^2$ ]{\includegraphics[width=0.3\textwidth]{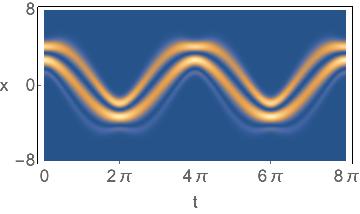}}

\vskip1ex
\subfigure[$(3,1)$, $\vert \psi_0 (x,t)\vert^2$   ]{\includegraphics[width=0.3\textwidth]{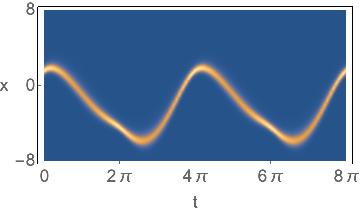}}
\hskip1ex
\subfigure[$(3,1)$, $\vert \psi_1 (x,t)\vert^2$  ]{\includegraphics[width=0.3\textwidth]{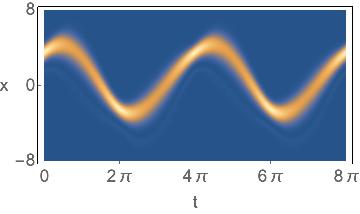}}
\hskip1ex
\subfigure[ $(3,1)$, $\vert \psi_2 (x,t)\vert^2$ ]{\includegraphics[width=0.3\textwidth]{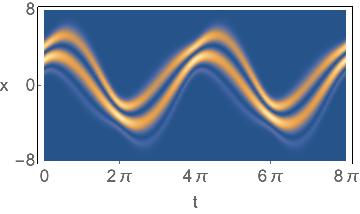}}

\caption{\footnotesize 
Probability densities of the three first Darboux deformed wave-packets $\psi_n(x,t)$ associated with the potentials shown in Figure~\ref{figpot} and Figure~\ref{PotV1}.
}
\label{DWP}
\end{figure}

The behavior of the three first states $\psi_n(x,t)$ of the nonstationary oscillators (\ref{V1-Ex2}) is exhibited in Figure~\ref{DWP} for the same parameters as the Figures~\ref{figpot} and \ref{PotV1}. The value $\varepsilon = -\frac12$ permitted the construction of the missing state $\psi_0(x,t)$. Notice that the maximum of such packet follows the time-dependent perturbation of the related potential, as expected. The same global behavior is appreciated for the wave-packets $\psi_1(x,t)$ and $\psi_2(x,t)$, where their local maxima follow the potential perturbation as the time goes on.

\begin{figure}[htb]

\centering
\subfigure[$(0,0)$, $z=i$ ]{\includegraphics[width=0.3\textwidth]{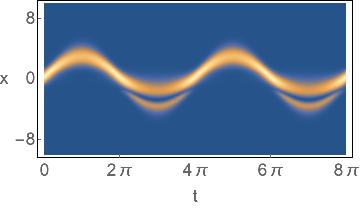}}
\hskip1ex
\subfigure[$(3,0)$, $z=i$ ]{\includegraphics[width=0.3\textwidth]{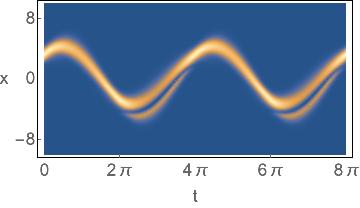}}
\hskip1ex
\subfigure[ $(3,1)$, $z=i$ ]{\includegraphics[width=0.3\textwidth]{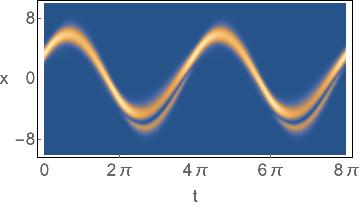}}

\vskip1ex
\subfigure[$(0,0)$, $z = 3-3i$ ]{\includegraphics[width=0.3\textwidth]{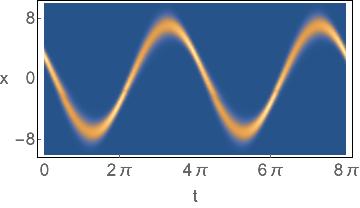}}
\hskip1ex
\subfigure[$(3,0)$, $z = 3-3i$ ]{\includegraphics[width=0.3\textwidth]{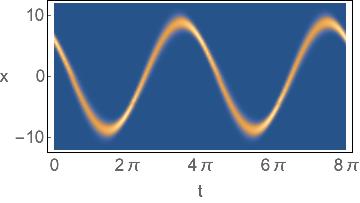}}
\hskip1ex
\subfigure[ $(3,1)$, $z = 3-3i$ ]{\includegraphics[width=0.3\textwidth]{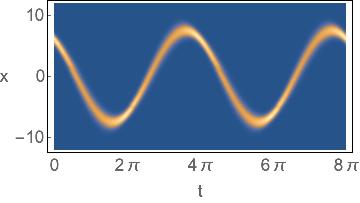}}

\caption{\footnotesize 
Probability density of the coherent states $\psi_z(x,t)$ for the indicated values of the initial point $(\langle \hat x \rangle_0, \langle \hat p \rangle_0)$ and the eigenvalue $z$. The other parameters are the same as  those of Figure~\ref{figpot}.
}
\label{figCS}
\end{figure}

 The coherent states $\psi_z(x,t)$ for the potential (\ref{V1-Ex2}) are shown in Figure~\ref{figCS} for the same parameters as in the previous figures and two different values of the complex eigenvalue $z$. The case with $z=i$ (upper row in the figure) exhibits the propagation of two maxima that obey their presence to the logarithmic derivative of $F(\chi)$ in the first term of $\psi_z(x,t)$, see Eq.~(\ref{cs2}). Such an effect becomes negligible for other values of $z$, as it can be noted in the plots of the lower row.

\subsection{Case $\varepsilon = -\frac32$}

\begin{figure}[htb]

\centering
\subfigure[$(0,0)$  ]{\includegraphics[width=0.3\textwidth]{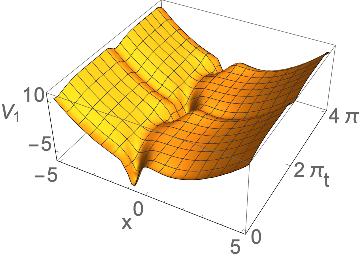}}
\hskip1ex
\subfigure[$(3,0)$ ]{\includegraphics[width=0.3\textwidth]{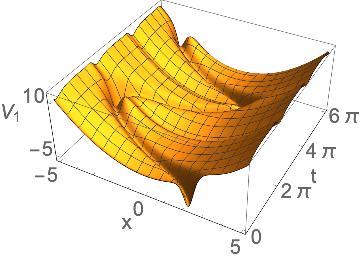}}
\hskip1ex
\subfigure[ $(3,1)$ ]{\includegraphics[width=0.3\textwidth]{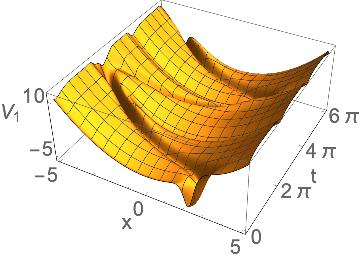}}

\caption{\footnotesize 
Nonstationary oscillators $V_1(x,t)$ defined in Eq.~(\ref{V1-Ex1}) for the indicated values of the initial point $(\langle \hat x \rangle_0, \langle \hat p \rangle_0)$. In all cases $t_0=0$, $m=1$, $\omega_0 =0.5$, and $k_a = 1.7k_b$. We have used $\alpha(t)$ with $a=1$ and $c=5$.
}
\label{PotV12}
\end{figure}

\begin{figure}[htb]

\centering
\subfigure[$(0,0)$, $\vert \psi_0 (x,t)\vert^2$  ]{\includegraphics[width=0.3\textwidth]{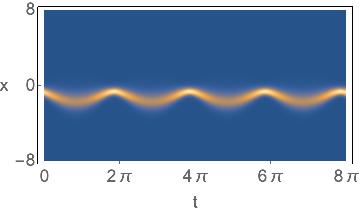}}
\hskip1ex
\subfigure[$(0,0)$, $\vert \psi_1 (x,t)\vert^2$ ]{\includegraphics[width=0.3\textwidth]{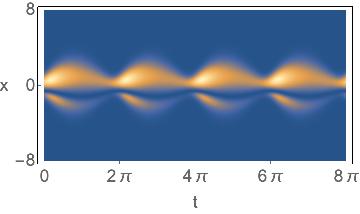}}
\hskip1ex
\subfigure[ $(0,0)$, $\vert \psi_2 (x,t)\vert^2$  ]{\includegraphics[width=0.3\textwidth]{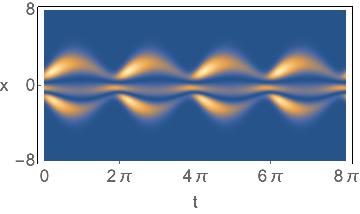}}

\vskip1ex
\subfigure[$(3,0)$, $\vert \psi_0 (x,t)\vert^2$  ]{\includegraphics[width=0.3\textwidth]{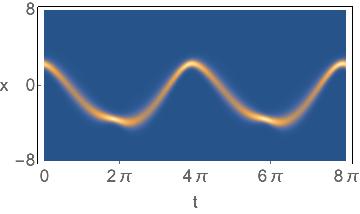}}
\hskip1ex
\subfigure[$(3,0)$, $\vert \psi_1 (x,t)\vert^2$ ]{\includegraphics[width=0.3\textwidth]{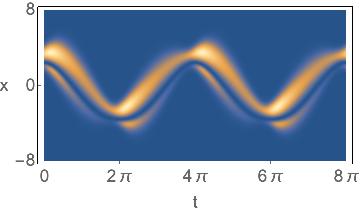}}
\hskip1ex
\subfigure[ $(3,0)$, $\vert \psi_2 (x,t)\vert^2$ ]{\includegraphics[width=0.3\textwidth]{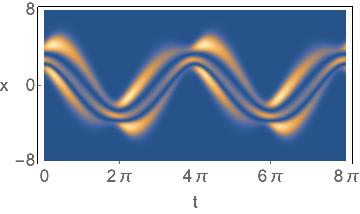}}

\vskip1ex
\subfigure[$(3,1)$, $\vert \psi_0 (x,t)\vert^2$   ]{\includegraphics[width=0.3\textwidth]{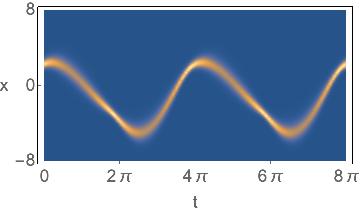}}
\hskip1ex
\subfigure[$(3,1)$, $\vert \psi_1 (x,t)\vert^2$  ]{\includegraphics[width=0.3\textwidth]{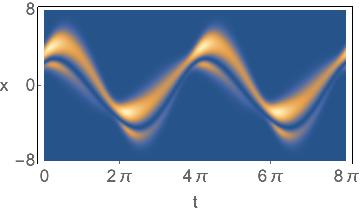}}
\hskip1ex
\subfigure[ $(3,1)$, $\vert \psi_2 (x,t)\vert^2$ ]{\includegraphics[width=0.3\textwidth]{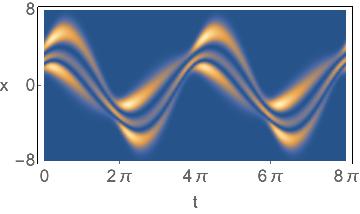}}

\caption{\footnotesize 
Probability densities of the three first Darboux deformed wave-packets $\psi_n(x,t)$ associated to the potentials shown in Figure~\ref{PotV12}.
}
\label{DWP2}
\end{figure}

\begin{figure}[htb]

\centering
\subfigure[$(0,0)$, $z=i$ ]{\includegraphics[width=0.3\textwidth]{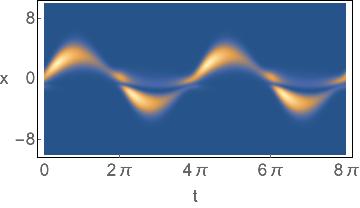}}
\hskip1ex
\subfigure[$(3,0)$, $z=i$ ]{\includegraphics[width=0.3\textwidth]{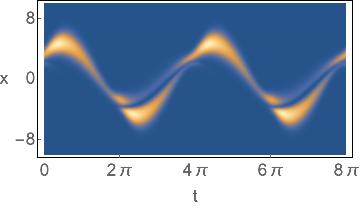}}
\hskip1ex
\subfigure[ $(3,1)$, $z=i$ ]{\includegraphics[width=0.3\textwidth]{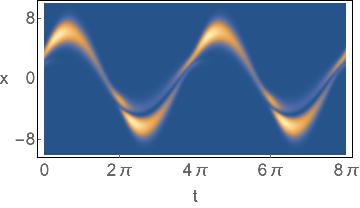}}

\vskip1ex
\subfigure[$(0,0)$, $z = 3-3i$ ]{\includegraphics[width=0.3\textwidth]{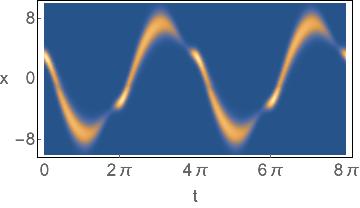}}
\hskip1ex
\subfigure[$(3,0)$, $z = 3-3i$ ]{\includegraphics[width=0.3\textwidth]{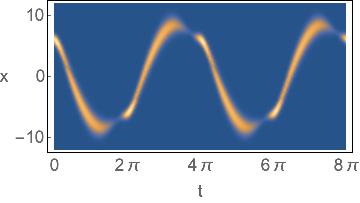}}
\hskip1ex
\subfigure[ $(3,1)$, $z = 3-3i$ ]{\includegraphics[width=0.3\textwidth]{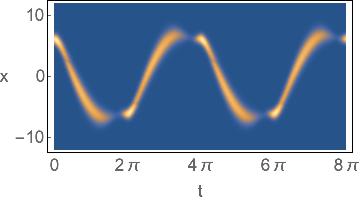}}

\caption{\footnotesize 
Probability density of the coherent states $\psi_z(x,t)$ for the indicated values of the initial point $(\langle \hat x \rangle_0, \langle \hat p \rangle_0)$ and the eigenvalue $z$. The other parameters are the same as  those of Figure~\ref{PotV12}.
}
\label{figCS2}
\end{figure}

In this case, the relevant results are the $F$-function
\be
\label{F-Ex2}
F(\chi) = k_a + \chi e^{\chi^2} \left[k_b + \sqrt{\pi} k_a \; \mathrm{Erf}(\chi) \right],
\ee
as well as the operator $\hat L$ and the potential $V_1(x,t)$ given by
\begin{subequations}
\be
\label{L-Ex1}
L = - 2\left(\frac{2m\omega_0}\hbar\right)^{1/2} \left[\frac{2k_a \chi e^{-\chi^2} + k_b (1 + 2 \chi^2) + \sqrt{\pi} k_a \mathrm{Erf}(\chi)}{k_a e^{-\chi^2} + \chi \left[k_b + \sqrt{\pi} k_a \mathrm{Erf}(\chi)\right]}\right]
- 2 e^{-i\theta(t)} A^-
\ee
and
\be
\label{V1-Ex1}
V_1(x) = \frac{m \omega_0^2 x^2}{2} - \frac{2\hbar^2}{\alpha(t)} \left(\frac{2m\omega_0}{m \hbar}\right)^{1/2} \partial_x 
\left[\frac{2k_a \chi e^{-\chi^2} + k_b (1 + 2 \chi^2) + \sqrt{\pi} k_a \mathrm{Erf}(\chi)}{k_a e^{-\chi^2} + \chi \left[k_b + \sqrt{\pi} k_a \mathrm{Erf}(\chi)\right]} \right]  + \frac{2\hbar \omega_0}{\alpha^2(t)}.
\ee
\end{subequations}
The time-evolution of the nonstationary potential (\ref{V1-Ex1}) is shown in Figure~\ref{PotV12} with similar values as those used in the previous case. Notice that the global behavior of an oscillating perturbation is also presented in this case. Similar conclusions are obtained from Figures~\ref{DWP2} and  \ref{figCS2}, where we show the time-evolution of the probability densities of the wave-packets $\psi_n(x,t)$ and coherent states $\psi_z(x,t)$, respectively.

\section{Conclusions}
\label{conclu}

We have constructed wave-packets with the Hermite-Gauss profile for the stationary oscillator of constant frequency $\omega_0$. These states are not eigenfunctions of the related Hamiltonian and are not orthonormal if the elements in the product are evaluated at different times. Nevertheless, we have shown that there exists an invariant operator $\hat I(t)$ which admits the Hermite-Gauss modes as eigenfunctions. Then, such functions form an orthonormal basis for the space of states of the stationary oscillator and differ from the solutions of the related Schr\"odinger equation just by a time-dependent phase.

The Hermite-Gauss modes have been used to construct time-dependent Darboux deformations of the stationary oscillator via the method introduced in Refs.~\cite{Bag95a,Bag95b,Bag96a,Bag96b}. We have shown that the new nonstationary oscillators exhibit a local `deformation' that oscillates along the parabola defined by the potential of the stationary case. In turn, the solutions are such that their local maxima also oscillates by following the deformation of the potential as the time goes on.

We have provided the invariant $\hat I_G$ for the nonstationary oscillators, so that the solutions reported here are eigenfunctions of $\hat I_G$ since the corresponding Hamiltonian is not an integral of motion of the system. The invariant operator $\hat I_G$ coincides with the invariant $\hat I$ of the Hermite-Gauss modes when the time-dependence of the nonstationary oscillators (represented by an additive operator $\hat G$ in the new Hamiltonian) is turned off.

We also provided the dynamical algebras for both sets of functions, the Gauss-Hermite modes and the solutions to the nonstationary oscillators, and show that they close the Heisenberg algebra. Then we have constructed the corresponding coherent states, which form an overcomplete set while they minimize the quadratures associated with the ladder operators. Remarkably, these states are not time-invariant since their time-dependence does not lies in the complex eigenvalue $z$, but on the basis of solutions itself.

It is expected that our approach can be applied to study either trapping of particles by electromagnetic fields \cite{Pau90,Gla92,Rosu96,Mih09,Mih18,Mih19}, or the propagation of electromagnetic signals \cite{Cru17,Gre17,Gre19}. Our approach can be extended to the case of non-Hermitian Hamiltonians \cite{Ros15,Bla18}, for which some interesting results have been reported quite recently \cite{Cen19}. Immediate applications are available in supersymmetric quantum mechanics \cite{Mie04,And04,Kha04,Suk04}, which can be used to model photonic systems with variable refractive index \cite{Man08,Con19a, Raz19,CruT15a,CruT15b}.

\appendix
\section{Invariant operator}
\label{ApA}

\renewcommand{\thesection}{A-\arabic{section}}
\setcounter{section}{0}  

\renewcommand{\theequation}{A-\arabic{equation}}
\setcounter{equation}{0}  

Any invariant operator (first integral) $\hat I(t)$ must satisfy the Heisenberg equation
\be
\frac{d}{dt} \hat I(t)=\frac{i}{\hbar}[ \hat H(t),I(t)]+\frac{\partial}{\partial t} \hat I(t)=0.
\label{eq:INV1}
\ee
For time-dependent oscillators the operator $\hat I (t)$ was achieved in mathematical form by Ermakov \cite{Erm80}. Fundamental results addressed to face nonstationary systems in quantum mechanics were then reported by Lewis and Riesenfeld \cite{Lew68,Lew69}, and formalized by Dodonov and Man'ko \cite{Dod75,Dod89}, and by Glauber \cite{Gla92}. Over the time, some approaches have been developed  to study a wide diversity of quantum mechanical problems (see, e.g. \cite{Cas13,Sch13,Cru15,Cru16,Dod00a,Man14,Gue15,Pad18,Gal18}), including the application of the Ermakov equation in coordinate representation (rather than using the time parameter) to construct stationary non-Hermitian exactly solvable Hamitlonians  \cite{Ros15,Bla18}. Recent results show that the invariant $\hat I(t)$ is a natural consequence of point transformations when nonstationary oscillators are produced as deformations of the stationary case \cite{Zel19a,Zel19b}. The relevance of $\hat I(t)$ is that one can find a set of its eigenfunctions
\be
\hat I(t) \overline{\varphi}_{n}(x,t)=\lambda_{n}\overline{\varphi}_{n}(x,t), \quad \lambda_n \not= \lambda_n (t), \quad n \in {\cal I} \subset \mathbb Z,
\label{eq:INV2}
\ee
which satisfy an orthonormality condition when the involved functions are evaluated at the same time. Thus, the product between $\overline{\varphi}_n (x,t)$ and $\overline{\varphi}_m (x,t')$ is not necessarily $\delta_{n,m}$ if $t\neq t'$. The latter is relevant since nonstationary Hamiltonians $\hat H(t)$ are not integrals of motion for the related system, so that the spectral problem defined by $\hat H(t)$ is either cumbersome or even intractable. In general, the eigenfunctions $\overline{\varphi}_n (x,t)$ of the invariant $\hat I(t)$ are connected with the solutions of the Schr\"odinger equation $i\hbar\partial_{t}\varphi_{n}(x,t)= \hat H(t)\varphi_{n}(x,t)$ through a time-dependent complex phase \cite{Lew69}. Namely, $\varphi_{n}(x,t) = e^{i\theta_{n}(t)}\overline{\varphi}_{n}(x,t)$, with $\theta_{n}(t)$ to be determined.

To obtain the invariant operator $\hat I_G$ reported in (\ref{invariante2}) we pay attention to the Hamiltonian defined by the time-dependent potential (\ref{vtemp}):
\[
V_1(x,t) = \frac12 m \omega_0^2 x^2 - \frac{\hbar^2}{m} \partial_x^2 \left[ \ln F(\chi(x,t))  \right] + \frac{2 \hbar \omega_0}{\alpha^2(t)}.
\]
That is, we use the Hamiltonian
\be
\hat H_{1}(t)=\frac{\hat p^{2}}{2m}+\frac{1}{2}m \omega_0^2 \hat x^2 - \frac{\hbar^2}{m} \hat G(\hat x, t), 
\label{eq:INV4}
\ee
where the operator $\hat G( \hat x, t)$ is defined such that
\be
\langle x \vert \hat G(\hat x, t) \vert x \rangle: = \frac{\hbar^2}{m} \partial_x^2 \left[ \ln F(\chi(x,t))  \right] - \frac{2 \hbar \omega_0}{\alpha^2(t)}.
\label{eq:INV5}
\ee
From the well known structure of the Ermakov-Lewis-Riesenfeld invariant we now propose
\be
\hat I_G (t) = c_1(t) \hat x^2+ c_2 (t) \hat p^{2}+ c_3(t) \{ \hat x , \hat p \} + c_4 (t) \hat G(\hat x, t),
\label{eq:INV6}
\ee
where the coefficients $c_k(t)$ are to be determined. Equation (\ref{eq:INV1}) is easily achieved by considering the relationships
\be
[\hat x^2, \hat p^2]=-[\hat p^{2}, \hat x^2] = 2i \hbar\{ \hat x, \hat p \}, \quad [\hat x^2, \{ \hat x, \hat p \}] = 4i \hbar \hat x^2, \quad [ \hat p^2,\{ \hat x, \hat p \}] = -4i \hbar \hat p^2, 
\label{eq:INV7}
\ee
along with the identity
\be
[\{\hat x, \hat p \}, f( \hat x)]= 2 \hat x [\hat p , f (\hat x)],
\label{eq:INV8}
\ee
with $f(\hat x)$ a smooth function of $\hat x$. Besides, it is straightforward to show that
\be
\frac{\partial}{\partial t} \hat G( \hat x, t)=-\frac{2\dot{\alpha}}{\alpha}\hat G (\hat x, t) -\frac{i}{\hbar}\frac{\dot{\alpha}}{\alpha}\hat x [ \hat p, \hat G(\hat x, t)],\quad \dot{\alpha}= \frac{\partial\alpha}{\partial t} .
\label{eq:INV9}
\ee
Therefore,
\be
\begin{alignedat}{3}
&\frac{d}{dt} \hat I(t) = && \left(\frac{c_1}{m}-m \omega_0^2 c_2 + m\dot c_3 \right) \{\hat x, \hat p\} + \left( \frac{2c_3}{m}+\dot c_2 \right) \hat p^2 + \left( -2m \omega_0^2c_3+\dot c_1 \right) \hat x^2 \\
& && + \frac{i}{2m\hbar} \left(c_4+2\hbar^2c_2\right) [\hat p^2, \hat G( \hat x, t)] + \frac{i}{\hbar} \left( \frac{2\hbar^2}{m}c_3- \frac{\dot{\alpha}}{\alpha} c_4 \right) \hat x [ \hat p, \hat G( \hat x, t)]\\
& && + \left(\dot c_4 - 2 \frac{\dot{\alpha}}{\alpha}c_4 \right) \hat G(\hat x, t) =0
\end{alignedat}
\label{eq:INV10}
\ee
implies the set of equations
\be
\begin{aligned}
& c_1-m^2 \omega_0^2 c_2 + m \dot c_3=0 \, ,\quad 2c_3 + m \dot c_2=0 \, ,\quad -2m \omega_0^2c_3+\dot c_1 =0, \\
& c_4+2\hbar^2c_2=0, \quad \frac{2\hbar^2}{m}c_3 -\frac{\dot{\alpha}}{\alpha} c_4 =0 , \quad \dot c_4 - 2 \frac{\dot{\alpha}}{\alpha} c_4=0.
\end{aligned}
\label{eq:INV11}
\ee
The straightforward calculation gives
\be
\begin{aligned}
c_1 = \dot{\alpha}^2+ \frac{4 \omega_0^2}{\alpha^2}, \quad c_2=\frac{\alpha^2}{m^2}, \quad c_3 =-\frac{\alpha\dot{\alpha}}{m}, \quad c_4 =-\frac{2\hbar^2 \alpha^2}{m^2},
\end{aligned}
\label{eq:INV12}
\ee
where we have used the Ermakov equation (\ref{erma}). After introducing the above parameters into Eq.~(\ref{eq:INV6}) we recover the expression (\ref{invariante2}) for the invariant $\hat I_G$. Clearly, if the operator $\hat G(\hat x,t)$ is turned off, then $\hat I_{G=0} =\hat I$, with $\hat I$ the invariant reported in Eq.~(\ref{invariante1}). The latter is quite natural by considering that the operator $\hat G(\hat x,t)$ corresponds to the time-dependent term that results from the difference $V_1(x,t) - V_0(x)$. Thus, if  $\hat G(\hat x,t)=0$ one has $V_1(x,t) =V_0(x)$.

\section*{Acknowledgment}

This research was funded by Consejo Nacional de Ciencia y Tecnolog\'ia (Mexico), grant number A1-S-24569, and by Instituto Polit\'ecnico Nacional (Mexico), project SIP20195981. K. Zelaya acknowledges the support from the Laboratory of Mathematics Physics, Centre de Recherches Math\'ematiques, through a postdoctoral fellowship. R. Razo acknowledges the funding received through the CONACyT scholarship number 421572.


\end{document}